\documentclass[final,3p,times,twocolumn]{elsarticle}
\usepackage[pdftex]{hyperref}

\usepackage{amssymb}
\usepackage{amsmath}

\journal{Physica D}

\begin{document}

\begin{frontmatter}

\title{Two-dimensional patterns in dip coating - first steps on the continuation path}

\author[muenster]{Phong-Minh Timmy Ly \fnref{fn2}}%[orcid=0000-0003-3836-120X]
\ead{timmy.ly@wwu.de}
\fntext[fn2]{ORCID: 0000-0003-3836-120X} 
\author[muenster]{Kevin David Joachim Mitas}
\author[muenster,cenos]{Uwe Thiele \fnref{fn1}}%\fnref{fn1}}%[orcid=0000-0001-7989-9271]
\ead{u.thiele@uni-muenster.de}
\ead[url]{http://www.uwethiele.de}
\fntext[fn1]{ORCID: 0000-0001-7989-9271} 
\author[muenster,cenos,son]{Svetlana V. Gurevich \fnref{fn3}}%[orcid=0000-0002-5101-4686]
\fntext[fn3]{ORCID: 0000-0002-5101-4686} 
%\author[technion]{Ofer Manor}
%\author[funsom]{Lifeng Chi}

\address[muenster]{Institute for Theoretical Physics, University of M\"unster, Wilhelm-Klemm-Strasse 9, D-48149 M\"unster, Germany}
\address[cenos]{Center for Nonlinear Science (CeNoS), University of M\"unster, Corrensstrasse 2, D-48149, M\"unster, Germany}
\address[son]{Center for Soft Nanoscience (SON), University of M\"unster, Busso-Peusstrasse 10, D-48149, M\"unster, Germany}
% \address[cmtc]{Center for Multiscale Theory and Computation (CMTC), University of M\"unster, Corrensstr. 40, 48149 M\"unster, Germany}
%\address[technion]{Wolfson Department of Chemical Engineering, Technion-Israel Institute of Technology, Haifa, 32000, Israel}
%\address[funsom]{Institute of Functional Nano \& Soft Materials (FUNSOM) and Jiangsu Key Laboratory for Carbon-based
%	Functional Materials \& Devices Collaborative Innovation Center of Suzhou Nano Science and Technology,
%	Soochow University, 215123 Suzhou, P. R. China}

\begin{abstract}
  We present a brief comparative investigation of the bifurcation
  structure related to the formation of two-dimensional deposition
  patterns as described by continuum models of Cahn-Hilliard type. These are, on the one hand a driven Cahn-Hilliard model for Langmuir-Blodgett transfer of a
  surfactant layer from the surface of a bath onto a moving plate and on the other hand a driven thin-film equation modelling
  the surface acoustic wave-driven coating of a plate by a simple
  liquid. In both cases, we present selected two-dimensional steady states corresponding to deposition patterns and discuss the main structure of the corresponding bifurcation diagrams.
\end{abstract}


\end{frontmatter}

\section{Introduction} \label{sec:intro}

When a solid substrate is withdrawn from a simple or complex liquid often a homogeneous or patterned deposit is left behind on the substrate. Such a deposition process forms the core of most coating technologies \cite{Wils1982jem,KhKS1992ces,WeRu2004arfm}. Patterned deposition may also occur when three-phase contact lines recede from solids due to evaporation \cite{Deeg2000pre,HaLi2012acie,FrAT2012sm,Rout2013rpp,DeDG2016epje}. Coating processes are employed in many branches of technology as a simple method to protect, embellish, level, texture and/or functionalize various types of surfaces. They all involve the transfer of material from a reservoir onto the surface of a solid. Often, their aim is to produce homogeneous coatings, although increasingly, also patterned deposition is aimed at.

Paradigmatic examples are dip coating where a solid plate is withdrawn from a bath and a liquid meniscus (or front) that recedes along a plate either due to evaporation or due to an imposed pressure gradient. In both cases, the meniscus recedes in the frame of the plate.  Note that the (mean) withdrawal speed can be naturally emerging, e.g., due to evaporation, or be imposed by external means, e.g., by an imposed plate speed or pressure gradient. In Ref.~\cite{Thie2014acis}, one distinguishes these set-ups as passive and active, respectively. The active set-up is more versatile as the imposed (mean) withdrawal speed represents a natural control parameter. In contrast, for the passive set-up the (mean) withdrawal speed naturally emerges from the evaporation dynamics. The examples considered here belong to the active set-ups.

Depending on the properties of the liquid, wetting behaviour of the liquid on the substrate and withdrawal speed (or strength of evaporation), a large variety of homogeneous and patterned coatings can be produced. For instance, well controlled experiments with particle suspensions \cite{RDLL2006l,XuXL2007ace,BoDG2010l} or polymer solutions \cite{YaSh2005afm,HoXL2007am,KwYY2011sm,JEZT2018l} may produce regular line patterns orthogonal or parallel to the withdrawal direction \cite{XuXL2007ace,YaSh2005afm,JEZT2018l}. Other described patterns include interconnected or wavy stripes \cite{JEZT2018l}, ladder structures \cite{YaSh2005afm}, hierarchical line patterns \cite{FSLJ2002l}, droplet or hole arrays \cite{YaSh2005afm,HoXL2007am,JEZT2018l}, and branched structures \cite{GRDK2002jpspbp,PSMB2008prl}. Overviews are given in \cite{Thie2014acis,FrAT2012sm}. Note that systematic accounts of parameter dependencies of properties of deposition patterns and of the transitions between different patterns are rare \cite{RDLL2006l,JEZT2018l}.

Note that dip coating is closely related to the classical Landau-Levich problem \cite{LaLe1942apu}, i.e., the prediction of the withdrawal speed-dependent thickness of a homogeneously deposited liquid layer. This is experimentally and theoretically well studied for simple, partially wetting liquids in \cite{SZAF2008prl,DFSA2008jfm}  and \cite{SADF2007jfm,ZiSE2009epjt,GTLT2014prl,GLFD2016jfm,TWGT2019prf}, respectively. Modelling is often based on mesoscopic hydrodynamic thin-film theory \cite{OrDB1997rmp,CrMa2009rmp}. However, in most cases, only one-dimensional (1d) settings are modelled \cite{ZiSE2009epjt,GTLT2014prl,GLFD2016jfm,TWGT2019prf}. The behaviour is characterized in terms of bifurcation diagrams that show how different types of steady profiles (so-called meniscus states, foot states and Landau-Levich film states) \cite{ZiSE2009epjt,GTLT2014prl,GLFD2016jfm} and time-periodic behaviour (deposition of lines orthogonal to the withdrawal direction) \cite{TWGT2019prf} are related in parameter space. In Ref.~\cite{SADF2007jfm} a discussion of the relaxation of transversal perturbations of the moving contact line included. To our knowledge, systematic fully nonlinear two-dimensional (2d) considerations beyond a few time simulations \cite{WTGK2015mmnp,TWGT2019prf} do neither exist for simple nor complex liquids.
In a closely related system the meniscus is not driven by the withdrawal of the substrate plate but by surface acoustic waves (SAW) that propagate in the substrate \cite{AlMa2015pf,MoMa2017jfm}. The governing thin-film equations of SAW-driven coating and dip coating by plate withdrawal mainly differ in the driving term (e.g.~compare~\cite{GTLT2014prl} and \cite{MoMa2017jfm}). 

Another dip coating system is Langmuir-Blodgett (LB) transfer where a relatively dense layer of surfactant molecules is transferred from the surface of a liquid bath onto a moving plate \cite{Blod1935jacs}. For certain surfactant densities (surface pressures) and plate velocities, homogeneous transfer is unstable and various patterns  of surfactant deposition arise \cite{RiSp1992tsf,GlCF2000n,CHFC2005am}. The patterns consist of substrate regions where different surfactant phases are deposited, namely, liquid-expanded and liquid-condensed phases. The separation into these phases is caused by substrate-induced condensation \cite{RiSp1992tsf}, i.e., by an influence of the molecular interaction with the substrate on the surfactant phase transition. Most aspects of the process are successfully described by a generalized Cahn-Hilliard model for phase separation that entirely neglects hydrodynamic motion \cite{KGFT2012njp}. An extensive study in 1d employing numerical path continuation \cite{AllgowerGeorg1987} reveals an intricate bifurcation structure consisting of many branches of steady concentration profiles and time-periodic states interconnected by a plethora of local and global bifurcations \cite{KGFT2012njp,KoTh2014n} while time simulations in 2d show that 1d patterns are not always transversally stable \cite{KGFT2012njp,WiGu2014pre}. Again, to our knowledge no systematic 2d considerations exist beyond direct time simulations.

Here, we focus on SAW-driven coating and LB transfer as described by a thin-film model and a Cahn-Hilliard-type model, respectively, to perform a first bifurcation analysis of coating systems in a 2d geometry. Thereby, we employ numerical path continuation techniques \cite{DWCD2014ccp,EGUW2019springer} using the package \textsc{pde2path} \cite{UeWR2014nmma,Ueck2019ccp}. For recent examples and further details regarding the usage of continuation techniques for thin-film equations and Cahn-Hilliard equations see \cite{LRTT2016pf,EWGT2016prf,EGUW2019springer}.

The structure of our work is as follows.  First, both models and the employed numerical path continuation method are briefly introduced in section~\ref{sec:theory}. Then, sections~\ref{sec:results-lb} and~\ref{sec:results-saw} present the bifurcation diagrams and a selection of steady state profiles for LB transfer and SAW-driven menisci, respectively. Finally, section~\ref{sec:conclusion} discusses similarities and differences of the two systems, and gives a conclusion and outlook.

\section{Model equations and numerical treatment}\label{sec:theory}
\subsection{General form}

The here considered transport and balance equations describe the dynamics of scalar field variables like densities, concentrations and film heights through partial differential equations. For a single field, the dynamics may combine contributions of a mass-conserving flux $\mathbf{j}_c$ and a nonmass-conserving flux (or rate) $j_{\mathrm{nc}}$. For a scalar field $\phi(\mathbf{x},t)$ the kinetic equation (sometimes called ``evolution equation'') takes the form
\begin{equation}
  \partial_t \phi = -\nabla\cdot\mathbf{j}_c + j_{\mathrm{nc}} \;,
  \label{eq:fluxes}
\end{equation}
i.e., for $ j_{\mathrm{nc}}=0$ the equation reduces to a continuity equation.

The fluxes can be further split into contributions representing gradient dynamics and nongradient dynamics. The former can be written in terms of the variational derivative $\delta \mathcal{F} / \delta\phi$ (e.g., a chemical potential or a pressure) of an underlying energy functional $\mathcal{F}[\phi]$. In the latter case this is not possible. A general form of (\ref{eq:fluxes}) proposed in Ref.~\cite{EGUW2019springer} is

\begin{eqnarray}
  \partial_t\phi =&& \nabla\cdot\left[Q_c\nabla\left(\frac{\delta\mathcal{F}[\phi]}{\delta\phi} + \mu_c^{\mathrm{ng}}\right) - \mathbf{j}_c^{\mathrm{ng}}\right] \nonumber\\
  &&- \left(Q_{\mathrm{nc}}\frac{\delta\mathcal{F}[\phi]}{\delta\phi} + \mu_{\mathrm{nc}}^{\mathrm{ng}}\right).
\label{eq:general_eq}
\end{eqnarray}
Here, $\mu_c^{\mathrm{ng}}$ and $\mu_{\mathrm{nc}}^{\mathrm{ng}}$ are nonvariational chemical potentials, and $\mathbf{j}_c^{\mathrm{ng}}$ is a nonvariational mass-conserving flux. It includes all terms that cannot be written as the gradient of a chemical potential.
The introduction of Ref.~\cite{EGUW2019springer} gives an overview of common models that can be brought in the form (\ref{eq:general_eq}).

If the nonvariational contributions are all zero, Eq.~(\ref{eq:general_eq}) becomes a gradient dynamics with mass-conserving and nonmass-conserving contributions. The form may be derived from Onsager's variational principle, i.e., by minimising the Rayleighian with
respect to appropriate fluxes \cite{Doi2013}. This is briefly shown in the appendix of Ref.~\cite{Thie2018csa}. Then Eq.~\eqref{eq:general_eq} describes a dynamics that descends a gradient of the energy $\mathcal{F}$. Hereby, the mobility functions 
$Q_c$ and $Q_{\mathrm{nc}}$ are nonnegative.

Both models that we present in the following sections are mass-conserving, i.e., they correspond to  Eq.~(\ref{eq:general_eq}) without the nonmass-conserving term ($Q_{\mathrm{nc}} = 0$ and $\mu_{\mathrm{nc}}^{\mathrm{ng}} = 0$). Furthermore, all driving nongradient terms are of the form $\mathbf{j}_c^{\mathrm{ng}}=(g(\phi),0)$, i.e., there is no nongradient chemical potential ($\mu_c^{\mathrm{ng}} = 0$). The resulting general form of the here considered models is 
\begin{equation}
  \partial_t\phi = \nabla\cdot\left[Q_c(\phi)\nabla\left(\frac{\delta\mathcal{F}[\phi]}{\delta\phi}\right)
    -
\begin{pmatrix}
          g(\phi) \\
           0
         \end{pmatrix}
\right].
\label{eq:general_eq2}
\end{equation}
The specific considered models only differ in the choice of the energy functional $\mathcal{F}$, the specific form of the remaining mobility function $Q_c(\phi)$ and of the driving force into $x$-direction $g(\phi)$ that determines the nongradient flux.

\subsection{Langmuir-Blodgett transfer} \label{ssec:lb_theory}

In the case of Langmuir-Blodgett transfer the field variable $\phi$ is specified to be the order parameter field $c(\mathbf{x},t)$ that represents a scaled surfactant concentration that encodes the deviation from the value at the critical point, i.e., it can take negative values. The dynamical model is a driven Cahn-Hilliard equation obtained from Eq.~\eqref{eq:general_eq2} for $\phi=c$ when specifying the energy functional
\begin{equation}
	\mathcal{F}[c] = \int_\Omega \left[\frac{1}{2}|\nabla c|^2 + f(c)\right]\, \mathrm{d}\mathbf{x}
\label{eq:lb-energy}
      \end{equation}
to be the classical Cahn-Hilliard free energy \cite{CaHi1958jcp} where $f(c)$ is the double-well potential
\begin{equation}
	f(c) = -\frac{1}{2}c^2 + \frac{1}{4}c^4 \;.
\label{eq:lb-ff}
\end{equation}
and the mobility is a constant that is then absorbed into the time scale, i.e., $Q_c = 1$. $\Omega$ stands for the 1d or 2d substrate domain.
Without driving force ($g=0$), Eq.~\eqref{eq:general_eq2} with (\ref{eq:lb-energy}) and (\ref{eq:lb-ff}) then corresponds to the classical Cahn-Hilliard equation~\cite{Cahn1965jcp}.

In order to model Langmuir-Blodgett transfer of a surfactant layer from a bath onto a moving plate, two modifications are required. First, an advection term is included as a driving force that defines
a mass-conserving nongradient flux. 
\begin{equation}
	g = v c
\end{equation}
where $v$ is the plate velocity, here our main control parameter. Second, the local free energy $f(c)$ is amended by incorporating the effect of substrate-induced condensation. Namely, the influence of a space-dependent external field 
\begin{equation}
	\mu\zeta(\mathbf{x}) = -\frac{\mu}{2}\left[1 + \tanh\left(\frac{x-x_s}{l_s}\right)\right]
\end{equation}
is added to the double-well potential as $\mu\zeta c$. It tilts the potential when the surfactant layer has passed onto the plate and the molecular interaction between surfactant layer and substrate becomes effective. In consequence, the ordered dense liquid-condensed (LC) phase $c = 1$ is favored over the disordered less dense liquid-expanded (LE) phase $c = -1$ for $x>x_s$. The parameters $x_s$ and $l_s$ represent the position and width of the transition region, respectively. Their exact values are of minor importance. For details about the driven Cahn-Hilliard model and corresponding mesoscopic hydrodynamic models see \cite{KoGF2009el,KGFC2010l,KGFT2012njp,ThAP2016prf}

The finally resulting scaled nondimensional model writes
\begin{equation}
  \partial_t c(\mathbf{x},t) = -\nabla\cdot\left[\nabla\left(\Delta c + c - c^3 - \mu\zeta(\mathbf{x})\right) -
\begin{pmatrix}
          v c \\
           0
         \end{pmatrix}
  \right]\;.
	\label{eq:lb}
\end{equation}
In all calculations, we use Dirichlet boundary conditions on the bath side $\left(x = 0\right)$ that fix the concentration and keep the second derivative at zero. The other domain boundary is at $x = L_x$ and represents the plate far away from the bath. There, for simplicity, we set the first and third derivative to zero. In the case of a 2d domain $\Omega_2 = \left[0,L_x\right] \times [0,L_y]$, Neumann boundary conditions are applied in the $y-$direction. The complete set of boundary conditions is
\begin{align}
	&c\left(0,y\right) = c_0, \qquad \partial_{xx}c\left(0,y\right) = 0, \\
	&\partial_x c\left(L_x,y\right) = \partial_{xxx} c\left(L_x,y\right) = 0, \\
	&\partial_y c\left(x,0\right) = \partial_{yyy} c\left(x,0\right) = 0, \\
	&\partial_y c\left(x,L_y\right) = \partial_{yyy} c\left(x,L_y\right) = 0.
	\label{eq:lb_bc}
\end{align}
The numerical treatment is briefly discussed in Section~\ref{sec:numerics} while results are given in Section~\ref{sec:results-lb}.

\subsection{SAW-driven meniscus}\label{ssec:saw_theory}

In the case of SAW-driven coating the field variable $\phi$ is
specified to be the thickness profile $h(\mathbf{x},t)$ of the
free-surface liquid on the solid substrate in which the surface
acoustic wave (SAW) propagates. The dynamical model is a
dimensionless thin-film
equation obtained from Eq.~\eqref{eq:general_eq2} for $\phi=h$ when specifying the energy functional
\begin{equation}
 F[h] = \int_\Omega \left[\frac{1}{2\mathrm{We}_\mathrm{s}} (\nabla h)^2 + f(h) + \frac{G_0}{2} h^2 \right]\mathrm{d}\mathbf{x}
  \label{eq:saw-energy}
\end{equation}
to account for capillarity, wettability and gravity. Here
$\mathrm{We}_\mathrm{s}$ is the Weber number, encoding the ratio of
convective and capillary stress. The wetting potential for a partially wetting liquid is
\begin{equation}
 f(h) = \frac{\mathrm{Ha}}{h^2} \left(\frac{h_p^3}{5h^3}-\frac{1}{2}\right)
 \label{eq:saw-ff}
\end{equation}
where the nondimensional Hamaker constant $\mathrm{Ha}>0$ controls the equilibrium contact angle and $h_p$ the precursor film height. The parameter $G_0$ is the gravity number and the mobility is $Q_\mathrm{c} = h^3/3$ \cite{OrDB1997rmp,Thie2010jpcm}.  $\Omega$ stands for the 1d or 2d substrate domain.  For typical scalings used to obtain the nondimensional quantities see \cite{MoMa2017jfm,GTLT2014prl}.  Without driving force ($g=0$) and gravity, Eq.~\eqref{eq:general_eq2} with (\ref{eq:saw-energy}) and (\ref{eq:saw-ff}) then corresponds to the classical thin-film equation describing dewetting, for cases with different $f(h)$ see \cite{Genn1985rmp,BEIM2009rmp,StVe2009jpm}.

Here, the lateral driving term results from the SAW that transfer energy into the liquid and drive a directed flow \cite{MRFY2015pre,AlMa2015pf,AlMa2016pf}. Due to the high frequency of the induced Rayleigh surface acoustic wave one can average over the fast time scale of the acoustic waves and derive a thin-film equation with an effective lateral driving term acting on the time scale of the viscous flow in the film \cite{AlMa2015pf,MoMa2017jfm}. As a result an advection term is included as a driving force that defines a mass-conserving nongradient flux, namely a contribution to $g$ that is \begin{equation}
  U_\mathrm{s} = \frac{\epsilon_\mathrm{s}}{4} \,\frac{h \sinh(2h) - h \sin(2 h) + 2 \cos(h)\cosh(h)}{\cos(2h)+ \cosh(2h)}
 \label{sec:mod:saw:eq:saw}
\end{equation}
 where $\epsilon_\mathrm{s}$ is the SAW strength, here our main control parameter.
For completeness we also include the effect of a plate inclination
$\alpha$. This gives another contribution to $g$, namely
\begin{equation}
 U_\mathrm{g} = - \frac{G_0 \alpha}{3} h^3.
  \label{sec:mod:saw:eq:flux}
\end{equation}
The resulting dimensionless film thickness evolution equation is
\begin{equation}
  \partial_t h  = - \nabla \cdot\left[
\frac{h^3}{3} \nabla \bigg(\frac{1}{\mathrm{We}_\mathrm{s}} \Delta h
- f'(h) - G_0 h\bigg)
- \begin{pmatrix}
         U\\
           0
         \end{pmatrix}
\right]
\label{sec:mod:saw:eq:full}
\end{equation}
where the driving force in $x-$direction is $U=U_\mathrm{s}(h)+U_\mathrm{g}(h)$.
Note that the capillarity contribution corresponds to the Laplace pressure while $f'(h)$ corresponds to the negative of the Derjaguin (or disjoining) pressure. 
With the exception of the driving term (\ref{sec:mod:saw:eq:saw}), Eq.~(\ref{sec:mod:saw:eq:full}) is identical to thin-film equations employed for the description of dip coating (or the Landau-Levich problems). See Ref.~\cite{TWGT2019prf} and references therein. Therefore we expect the results presented here to be relevant for the Landau-Levich system as well.

% Therefore, the substrate is withdrawn.
% In more commonly systems instead, a plate is withdrawn with a velocity from a liquid bath reservoir.
%
%
%
We consider a geometry where the bath/meniscus is at $x=0$ and the SAW propagates in positive $x$-direction. The resulting boundary conditions for the 1d case with the domain $\Omega_1 = [0,L_{x}]$ are \cite{MoMa2017jfm}
\begin{align}
  h \vert_{x=0} &= h_m\,, \qquad \partial_{xx} h \vert_{x=0} = 1 \quad \text{and}
  \label{sec:mod:saw:eq:1d_bc_1}\\
  \partial_x h\vert_{x=L_{x}} &=\partial_{xx} h\vert_{x=L_{x}} =0\,. \label{sec:mod:saw:eq:1d_bc_2}
\end{align}
In the 2d case with the domain $\Omega_2 = [0,L_{x}]\times
[0,L_y]$ we additionally use
Neumann boundary conditions in lateral direction, i.e.,
\begin{equation}
 \partial_y h\vert_{y=0} =\partial_{yyy} h\vert_{y=0} = \partial_y h\vert_{y=L_{y}} =\partial_{yyy} h\vert_{y=L_{y}} =0\,.
 \label{sec:mod:saw:eq:2d_bc_y}
\end{equation}
The numerical treatment is briefly discussed in Section~\ref{sec:numerics} while results are given in Section~\ref{sec:results-saw}.
\begin{figure*}[tb]
	\centering
	\includegraphics[width=\hsize]{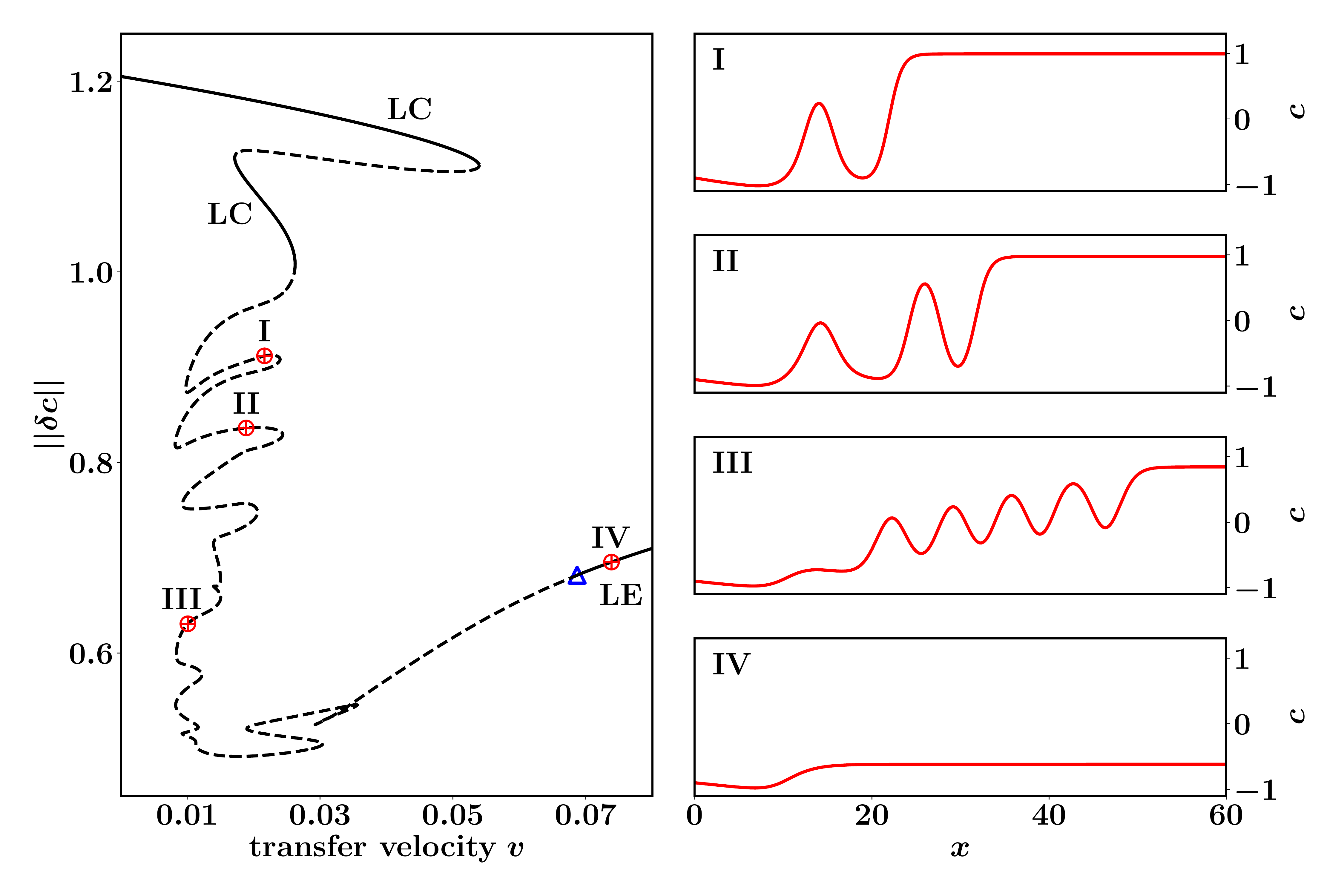}
	\caption{(left) Shown is for the case of Langmuir-Blodgett transfer the bifurcation diagram of 1d steady states as described by Eq.~(\ref{eq:lb}) in terms of the $L^2$ norm as a function of the transfer velocity $v$ (for a plate drawn to the left $v>0$). Linearly stable and unstable states are indicated by solid and dashed lines, respectively. ``LC'' and ``LE'' indicate on which stable sub-branches the liquid-condensed and liquid-expanded phase is homogeneously deposited. The red symbols and roman numbers indicate the loci of the concentration profiles shown on the right. The blue triangle indicates where the supercritical Hopf bifurcation occurs that stabilises the steady homogeneous transfer of the LE phase. The parameters are $\mu = 1$, $x_s = 10$, $l_s = 2$, $L_x = 60$ and $c_0 = -0.9$ while the data set can be found at \cite{LMTG2020zen}. 
	}
	\label{fig:lb_1d}
\end{figure*}

\subsection{Numerical path continuation}\label{sec:numerics}

As we are interested in the bifurcation structure of transversally inhomogeneous deposition in coating problems we entirely focus on results obtained with numerical path continuation methods \cite{AllgowerGeorg1987,DWCD2014ccp,EGUW2019springer} that have to our knowledge not yet been applied to such a problem.  

Path continuation is a very convenient method to obtain steady and stationary states of physical systems described by ordinary and partial differential equations without the use of time simulation. The states are directly tracked when varying a control parameter, the so-called continuation parameter. This allows one to determine stable and unstable states steady and time-periodic states, to detect bifurcations and to follow the emerging branches of new states. Furthermore tracking bifurcation points in two-parameter continuations helps to obtain an overview of the resulting structure and dynamic behavior of the investigated system. Many examples exist where such methods are applied to (driven) Cahn-Hilliard and thin-film equations in 1d \cite{Thie2010jpcm,GTLT2014prl,KoTh2014n,LRTT2016pf,TWGT2019prf,TSJT2020preprint} and 2d \cite{BeTh2010sjads,BKHT2011pre,EWGT2016prf,EnTh2019el} settings. The methods are implemented in many numerical toolboxes \cite{DWCD2014ccp}, for the considered equations the most commonly used ones are \textsc{auto07p} \cite{DoKK1991ijbc} and \textsc{pde2path} \cite{UeWR2014nmma,Ueck2019ccp,EGUW2019springer}. All calculations presented here are performed with the latter.

In the following we shortly summarise how pseudo-arclength continuation works.
Let the right-hand side of the considered kinetic equation (PDE) be denoted as the operator $G[\phi,\lambda]$, which acts on a continuous field $\phi$ and depends on a control parameter $\lambda$. For stationary solutions one then has
\begin{equation}
	0 = G[\phi,\lambda]\;.
\end{equation}
In \textsc{pde2path}, spatial discretization is applied by means of the finite element method, turning above equation into a set of algebraic equations
\begin{equation}
	0 = \boldsymbol{G}[\mathbf{\boldsymbol{\phi}},\lambda].
\label{eq:cont_rhs}
\end{equation}
where $\boldsymbol{G}$ and $\boldsymbol{\phi}$ are arrays.

Continuation along a branch of steady solutions of Eq.~(\ref{eq:cont_rhs}) then proceeds via prediction and correction steps. The prediction advances in $\lambda$ based on a tangent construction and 'locks in' a new value for the arclength $s$ along the branch. The correction step is based on 
employing Newton's method to solve Eq.~(\ref{eq:cont_rhs}) at fixed $s$, i.e., the exact value of $\lambda$ becomes part of the solutions. After converging, the next prediction step is taken. By continuous iteration this results in branches of solutions. Employing the arclength in the correction step allows for the continuation of solution branches around saddle-node bifurcations.

The used method is called ``pseudo-arclength continuation'' as an approximation of the arclength is employed, whereby each change $\Delta s$ is determined by
\begin{equation}
	|\Delta\boldsymbol{\phi}|^2 + (\Delta\lambda)^2 = (\Delta s)^2 \;.
\label{eq:delta_s}
\end{equation}
With other words, then Eq.~\eqref{eq:cont_rhs} and \eqref{eq:delta_s} are solved together for the extended array $\boldsymbol{\tilde{\phi}} = (\boldsymbol{\phi},s)$. The package 
\textsc{pde2path} also provides routines to detect various bifurcations and to switch to emerging branches.

Here, we characterise the calculated branches by the $L^2$-norm of the solution, i.e., we use
\begin{equation}
  ||\delta\phi|| = \frac{1}{\Omega}\int_\Omega \left[
    \phi^2
  \right]\mathrm{d}\mathbf{x}.
\end{equation}
Next we discuss results for both systems, namely in Sec.~\ref{sec:results-lb} the LB transfer and in Sec.~\ref{sec:results-saw} the SAW-driven coating. 

\begin{figure*}[tb]
	\centering
	\includegraphics[width=0.95\hsize]{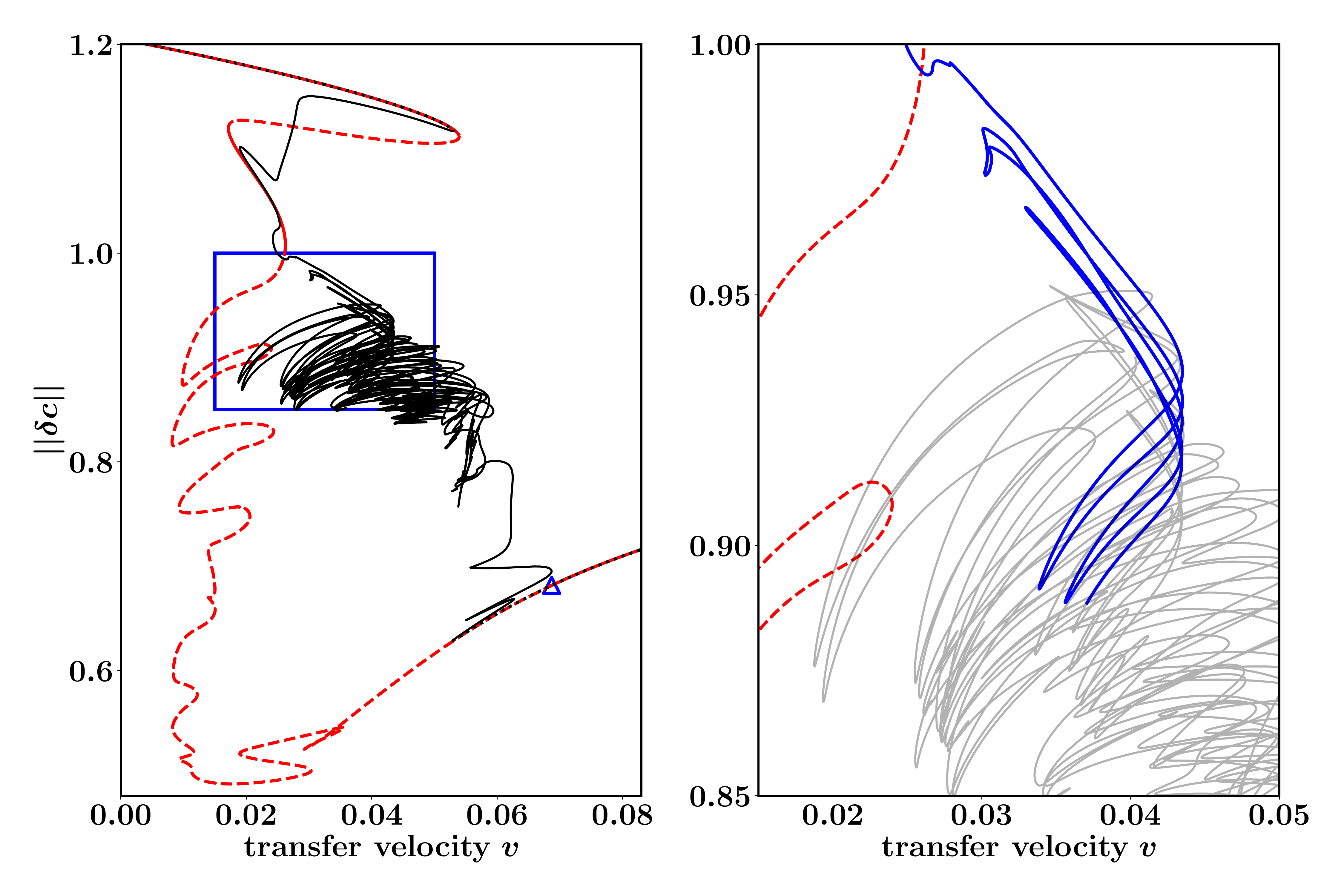}
	\caption{(left) Bifurcation diagram for the case of Langmuir-Blodgett transfer showing on the one hand the branch of 1d steady states of Fig.~\ref{fig:lb_1d} as red dashed (unstable) and solid (stable) lines and on the other hand a branch of fully 2d steady states as solid black line. Stability is solely indicated in the 1d case. The supercritical Hopf bifurcation in 1d almost coincides with the one in 2d and is indicated by a blue triangle. The blue box indicates the region whose zoom is given in the right panel. The close-up enlarges part of the intricate snaking curve. Thereby the first section of this curve is coloured blue to visually disentangle overlapping parts. The parameters are as in Fig.~\ref{fig:lb_1d} while the transversal size is $L_y = 40$. The data set can be found at \cite{LMTG2020zen}. Example 2d profiles are presented in Fig.~\ref{fig:lb_profiles} while a space-arclength plot is given in Fig.~\ref{fig:lb_space_arc}.
	}
	\label{fig:lb_2d}
\end{figure*}

\section{Results for Langmuir-Blodgett transfer}\label{sec:results-lb}

We first review information about the bifurcation structure of the Langmuir-Blodgett system obtained in the 1d case \cite{KGFT2012njp,KoTh2014n} (then with \textsc{auto07p}).
Here, \textsc{pde2path} is used, some parameters differ and downstream at $x = L_x$ the boundary condition $\partial_{xxx}c = 0$ is used instead of $\partial_{xx}c = 0$. This affects the results only quantitatively.

Figure~\ref{fig:lb_1d} shows a typical bifurcation diagram of Eq.~\eqref{eq:lb}-\eqref{eq:lb_bc} on the left and selected steady concentration profiles on the right. Solid and dashed lines represent linearly stable and unstable states while the control parameter is the transfer velocity $v$. The linearly stable sub-branch beginning at large $v$ represents homogeneous deposition of the LE phase whereas the remaining two linearly stable sub-branches at lower $v$ consist of homogeneous deposition of the LC phase. The profiles on the latter two only differ by the width of the low concentration region on the bath side (not shown in 1d, but cf.~panels (a) and (d) in Fig.~\ref{fig:lb_profiles} below). The parts of the branch where LE and LC phase are respectively deposited are connected by an irregularly snaking branch of unstable states. Passing along this snaking part, starting at small $v$ more and more undulations emerge at the front-like transition region between the low concentration of the bath and the deposited concentration. Furthermore, multiple Hopf bifurcations are present, many of them subcritical. Time-periodic states that emerge at these Hopf bifurcations  are mostly unstable and end at other Hopf bifurcations or at global bifurcations. For details see \cite{KGFT2012njp}. In Fig.~\ref{fig:lb_1d} we only mark by a blue triangle the most relevant Hopf bifurcation where a change in stability occurs. The branch of time-periodic states which emerges at this bifurcation (not shown) undergoes a secondary period-doubling bifurcation. The bifurcating branch consists of stable time-periodic solutions, that are obtained when performing time-simulations in the range between stable homogeneous deposition of LC and LE phases at respective small and large transfer velocities. The branch ends in a global bifurcation on one of the upper parts of the snaking branch of steady states. This is analysed in some detail in Ref.~\cite{KoTh2014n}.

\begin{figure*}[ph]
	\centering
	\includegraphics[width=0.8\hsize]{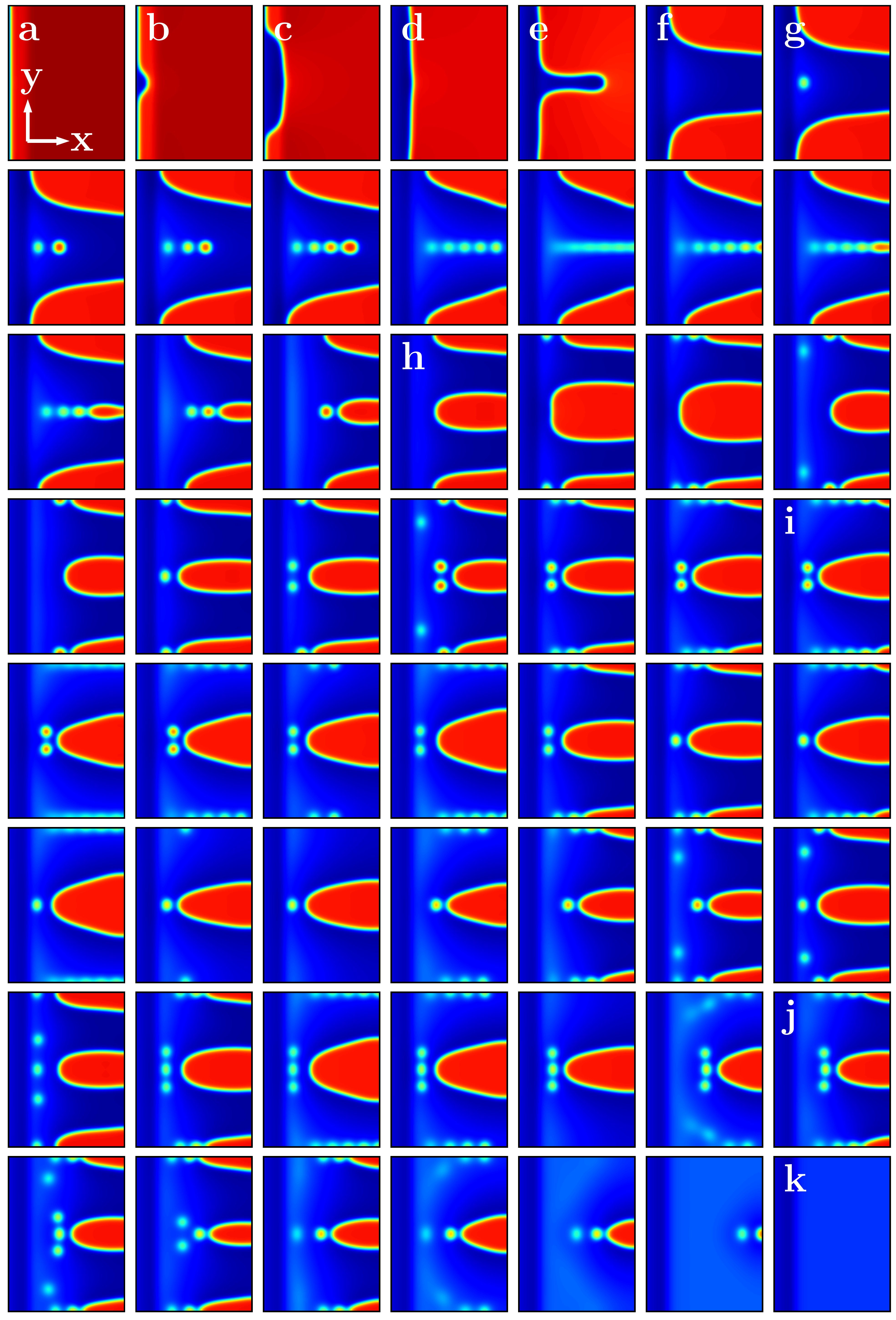}
	\caption{Shown are selected density profiles (for the corresponding colour bar see Fig.~\ref{fig:lb_space_arc}) of 2d steady states on the branch of fully 2d states in Fig.~\ref{fig:lb_2d}. From top left to bottom right we start at $v = 0$ first follow the branch of transversally invariant states, switch to the branch of 2d states and follow it through the end of Fig.~\ref{fig:lb_2d} (left), i.e.,  we follow the branch with increasing arclength $s$. All states except (a), (d) and (k) are unstable. Note that patterns are shown in the $y-$periodic $L_x\times 2L_y=60\times80$ domain. The remaining labels are used for reference in the main text. The remaining parameters are as in Fig.~\ref{fig:lb_2d}.
	}
	\label{fig:lb_profiles}
\end{figure*}

\begin{figure*}[ph]
	\centering
	\includegraphics[width=\hsize]{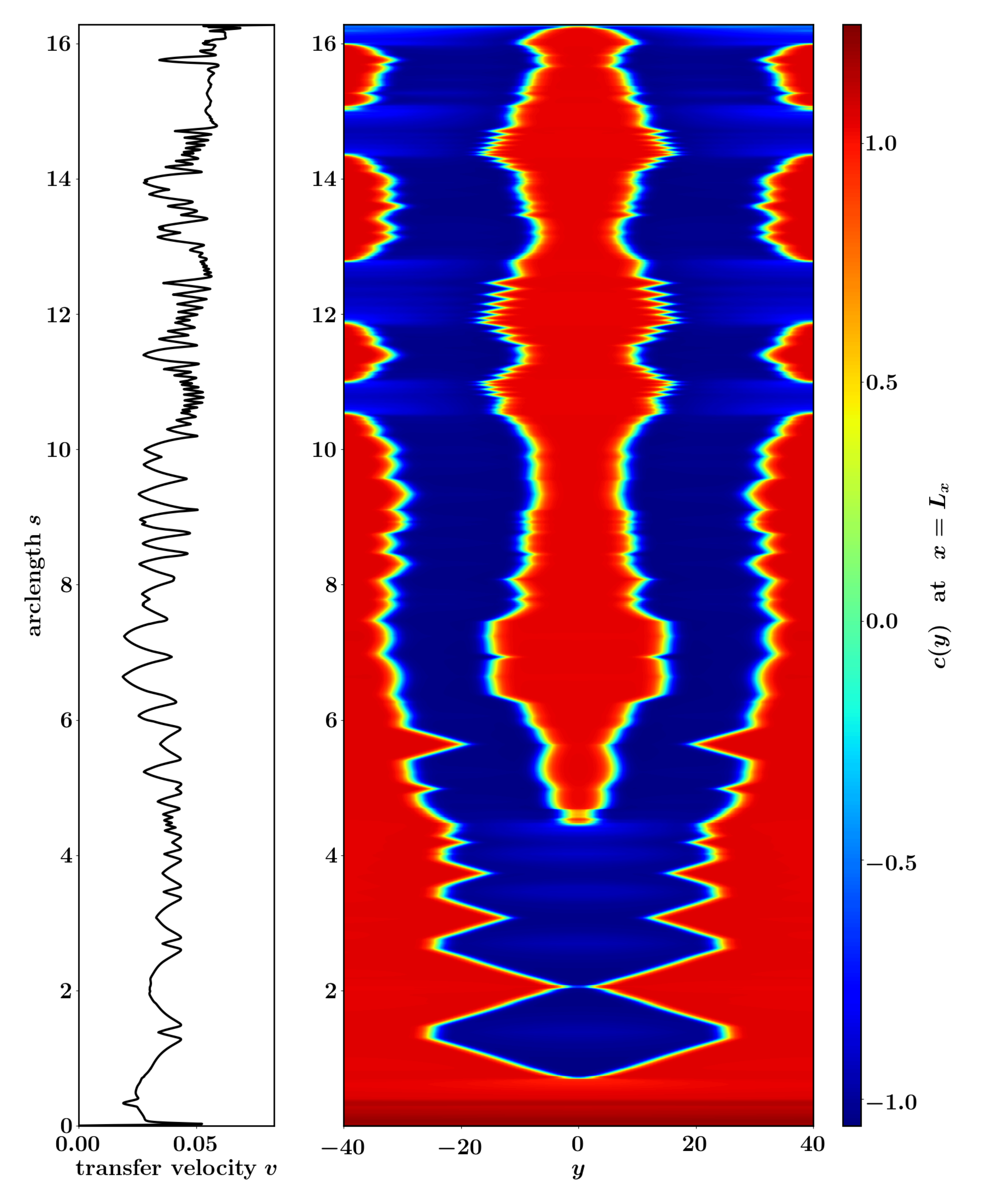}
	\caption{(right) Shown are space-arclength plots showing the concentration profile $c(x=L_x, y)$, i.e., on the downstream boundary, for all arclengths $s$ along the entire bifurcation curve of 2d states shown in Fig.~\ref{fig:lb_2d}. (left) The accompanying panel shows the corresponding values of the control parameter transfer velocity $v$. The remaining parameters are as in Fig.~\ref{fig:lb_2d}. 
	}
	\label{fig:lb_space_arc}
\end{figure*}
%\clearpage

Next, we transversally extend the domain and consider a fully 2d case with $L_x\times L_y = 60 \times 40$. The resulting bifurcation diagram is shown in Fig.~\ref{fig:lb_2d}(left) together with the 1d bifurcation curve already presented in Fig.~\ref{fig:lb_1d}. At small and large transfer velocities $v$ the 2d result (black dotted line) coincides with the 1d curve (red line), i.e., the solutions are translation invariant in transversal direction, shortly ``$y-$translation invariant''. Then, coming from small $v$, at about $v=0.053$ a pitchfork bifurcation occurs where two branches emerge subcritically (their respective norms coincide). They consist of states with broken $y-$translation symmetry. As we use Neumann boundary conditions (instead of periodic ones), the two emerging branches are related by a translation in $y$ by $L_y/2$. The emerging branch first extends towards smaller $v$, undergoes a small number of saddle-node bifurcations before it wildly folds back and forth producing a very dense cluster of sub-branches that overlay and cross each other multiple times. A zoom into part of this region is given in Fig.~\ref{fig:lb_2d}(right) where the first part of the branch is given in blue to allow one to distinguish some of the finer features. Solutions on the 2d branches are mostly unstable. Branches are linearly stable on the parts that coincide with stable parts of the 1d branch. Naturally, the black dotted part at high velocities gains stability only after the Hopf bifurcation which occurs at almost the same velocity as in 1d.

Fig.~\ref{fig:lb_profiles} presents a small selection of 2d density profiles along the whole solution branch shown in Fig.~\ref{fig:lb_2d}. Here, blue and red represent the low-concentration LE phase and the high-concentration LC phase, respectively. Note that calculations are performed using the $L_x\times L_y$ computational domain, while the images show a complete transversal period using the reflection symmetry at $y = 0$, i.e., they show a  $L_x\times 2L_y$ domain. In the sequence of the images we first follow the branch of transversally invariant states [panel (a)], switch to the branch of 2d states and follow it [panels (b) and (c)] through to its first fold at $v\approx 0.019$. It actually represents an imperfect pitchfork bifurcation as the state becomes again nearly $y-$translation invariant [panel (d)]. The 2d states between panels (a) and (d) can be seen as the emergence of the second stable $y-$translation invariant state via 2d states that represent 'nucleation' of a second 'terasse' in concentration (central blue bump in (a)) that then extends laterally, i.e., from (b) via (c) to (d) two red-to-blue (LC-to-LE) fronts move outwards. Beyond (d), one passes another imperfect pitchfork bifurcation, and the states become again truly two-dimensional. We see a central finger of blue prolong to the right [panels (e)] till it meets the right boundary. Caused by the finite system size one is then able to move between different branches that for a system semi-infinite in $y-$direction would all remain disconnected.

The different interconnected 2d states combine one or two fingers (blue in red) with different numbers and arrangements of red spots. The branch in Fig.~\ref{fig:lb_2d} forms a rather wild tangle corresponding to a complicated structure with many saddle-node bifurcations, accompanied by the occasional Hopf bifurcation (not shown). A detailed study of the structure and its dependence on parameters cannot be provided here. To give an impression of the complexity we present in Fig.~\ref{fig:lb_space_arc} an alternative representation. Namely, the concentration profile $c(x=L_x, y)$ on the downstream boundary (the right boundary) in Fig.~\ref{fig:lb_profiles} is presented in dependence of the arclength $s$ along the bifurcation curve, i.e., in a space-arclength plot. The accompanying panel shows the corresponding values of $v$, i.e., allows one to identify at which $s$ saddle-node bifurcations occur. The combination of the two panels allows one to appreciate the type of changes occurring along the bifurcation curve. 

For instance, one clearly discerns the formation of the first complete blue stripe at $s \approx 0.7$. It then widens and narrows several times accompanied by the emergence of a central line of red spots (not seen in the particular slice shown in Fig.~\ref{fig:lb_space_arc}(right), cuts at a different $x-$position would need to be used). The red spots first appear one after another [Fig.~\ref{fig:lb_profiles}(g) and second row of images], then nucleate a central red structure at the boundary [second last image in second row]. This structure then grows from the boundary eating up spot after spot [third row of Fig.~\ref{fig:lb_profiles} till Fig.~\ref{fig:lb_profiles}(h)]. In this way the original single blue finger is split into two. The emergence of the central red structure at the boundary is seen at $s \approx 4.42$ in Fig.~\ref{fig:lb_space_arc}.

From this point onwards a varying number of spots emerges and vanishes again at the left hand side tips of either of the red (LC) finger-like domains. In addition, from Fig.~\ref{fig:lb_profiles}(i) to~(j) the red domains on the lateral boundaries sometimes recede from the left boundary, i.e., from the bath, passing through intermediate states in the form of multiple spots. Finally the blue (LC) domain advances everywhere till the branch of 2d states ends in the pitchfork bifurcation on the homogeneous blue (LE) solution [Fig.~\ref{fig:lb_profiles}(k)]. 
The intricate sequence of widening and narrowing, advancing and receding fingers is well appreciated in the pattern of the space-arclength plot [Fig.~\ref{fig:lb_space_arc}]. The three sections of the 2d branch where the finger on the lateral boundary breaks up into a line of spots are easily distinguished and occur in the ranges $s=\{[10.5,10.9],[11.8,12.7],[14.3,15.0]\}$. Next we analyse the SAW-driven meniscus following the same scheme as done here for LB transfer.

\begin{figure*}[tbh]
\centering
\includegraphics[width=0.9\hsize]{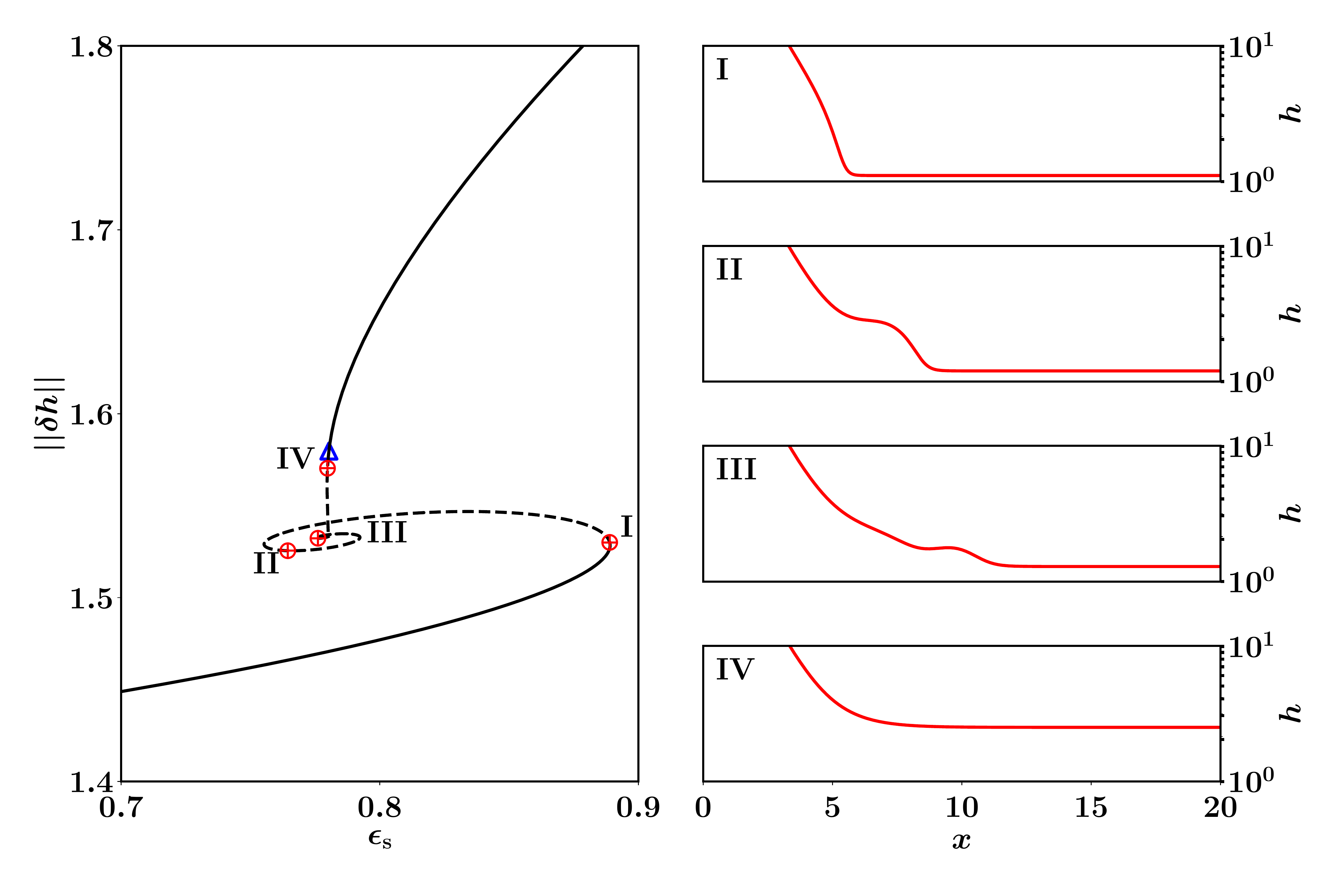}
\caption{(left) Shown is the bifurcation diagram for the case of a SAW-driven meniscus for 1d steady states as described by Eq.~(\ref{sec:mod:saw:eq:full}) in terms of the $L^2$ norm as a function of the SAW strength $\epsilon_\mathrm{s}$. Linearly stable and unstable states are indicated by solid and dashed lines, respectively. The red symbols and roman numbers indicate the loci of the concentration profiles shown on the right. The blue triangle indicates where the final Hopf bifurcation stabilises the Landau-Levich films. The parameters are $L_x = 40$, $h_m = 8.1$, $h_p =0.1$, $\mathrm{We}_\mathrm{s}= 2.0$, $\mathrm{Ha} =1.0$, $G = 10^{-3}$ and $\alpha = 0.2$. Shown is only the range $0\le x \le L_x/2$. The data set can be found at \cite{LMTG2020zen}. 
}
\label{sec:mod:saw:fig:one_dimensional}
\end{figure*}

\section{SAW-driven meniscus}\label{sec:results-saw}
We first determine the bifurcation structure of the SAW-driven liquid meniscus obtained for the steady version of Eq.~(\ref{sec:mod:saw:eq:full}) in the 1d case. The left panel of Fig.~\ref{sec:mod:saw:fig:one_dimensional} gives the bifurcation diagram in terms of the norm as a function of the SAW strength $\epsilon_\mathrm{s}$. On the right a number of corresponding steady thickness profiles is presented. Starting at small $\epsilon_\mathrm{s}$ one first encounters linearly stable meniscus solutions, i.e., the meniscus directly transitions into an ultrathin adsorption layer of height $h\approx h_p$ (profile~I). The branch undergoes a saddle-node bifurcation at $\epsilon_\mathrm{s}\approx0.88$ where it folds back and the solutions become unstable. Then the curve seems to spiral into a point at $\epsilon_\mathrm{s}\approx0.78$ becoming more and more unstable. Along this part of the branch the steady state develops a foot that is pushed out more and more by the SAW (profiles~II and~III). Then the branch becomes nearly vertical, more bifurcations
occur (not shown) that first destabilise and then stabilise the solutions. The steady state finally becomes stable at the final Hopf bifurcation, shown as a blue triangle in  Fig.~\ref{sec:mod:saw:fig:one_dimensional}, then tilts to the right. The profile becomes a meniscus that directly transitions into a Landau-Levich film that is pushed out of the meniscus. Beyond the range of the figure its thickness further increases with increasing $\epsilon_\mathrm{s}$, as captured by the power law $h_\mathrm{c}\sim \epsilon_\mathrm{s}^{2/3}$ that is exactly the same as in the classical Landau-Levich problem. 

Note that this behaviour is closely related to the one observed for the classical Landau-Levich system \cite{LaLe1942apu,SADF2007jfm,ZiSE2009epjt} when studied with a thin-film model incorporating wettability through a wetting energy \cite{GTLT2014prl,TWGT2019prf}. Namely it resembles a combination of scenarios (b) to (d) presented in Fig.~2 of Ref.~\cite{GTLT2014prl}. This hints at an approximately similar effect of dragging velocity $U$ in \cite{GTLT2014prl} as the SAW strength $\epsilon_\mathrm{s}$ controlled here. However, details regarding the changes in stability differ between the systems. This will be discussed elsewhere. A first bifurcation diagram for the fully 2d Landau-Levich system can be found in Fig.~14 of Ref.~\cite{EGUW2019springer}.

\begin{figure*}[tbh]
\includegraphics[width=\hsize]{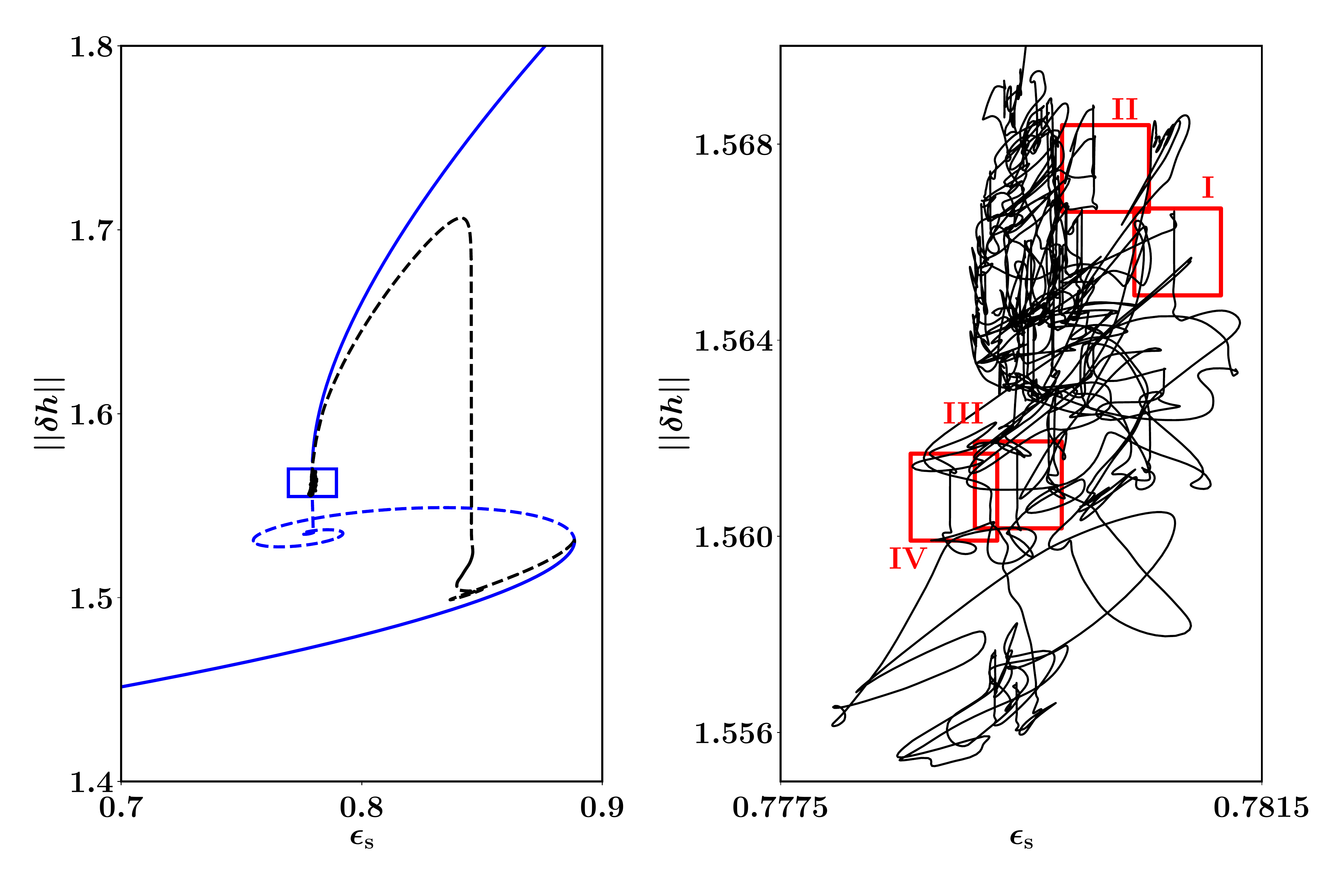}
\caption{(left) Bifurcation diagram for the case of a SAW-driven meniscus showing on the one hand the branch of 1d steady states of Fig.~\ref{fig:lb_1d} as blue dashed (unstable) and solid (stable) lines and on the other hand a branch of fully 2d steady states as solid (stable) and dashed (unstable) black line. The blue frame indicates the region whose zoom is given in the right panel. The close-up enlarges the barely visible intricately snaking part of the curve. The red frames highlight some regions with similar exponential snaking structures that are given in Fig.~\ref{sec:mod:saw:fig:two_dimensional_zoom_further}. The parameters are as in Fig.~\ref{sec:mod:saw:fig:one_dimensional} while the transversal size is $L_y = 15$. The data set can be found at \cite{LMTG2020zen}. Example 2d profiles are presented in Fig.~\ref{sec:mod:saw:fig:two_dimensional_steady_states} while a space-arclength plot is given in Fig.~\ref{sec:mod:saw:fig:two_dimensional_arclength_epsilon_y}.
}
\label{sec:mod:saw:fig:two_dimensional}
\end{figure*}

\begin{figure*}[tbh]
 \centering
\includegraphics[width=0.8\hsize]{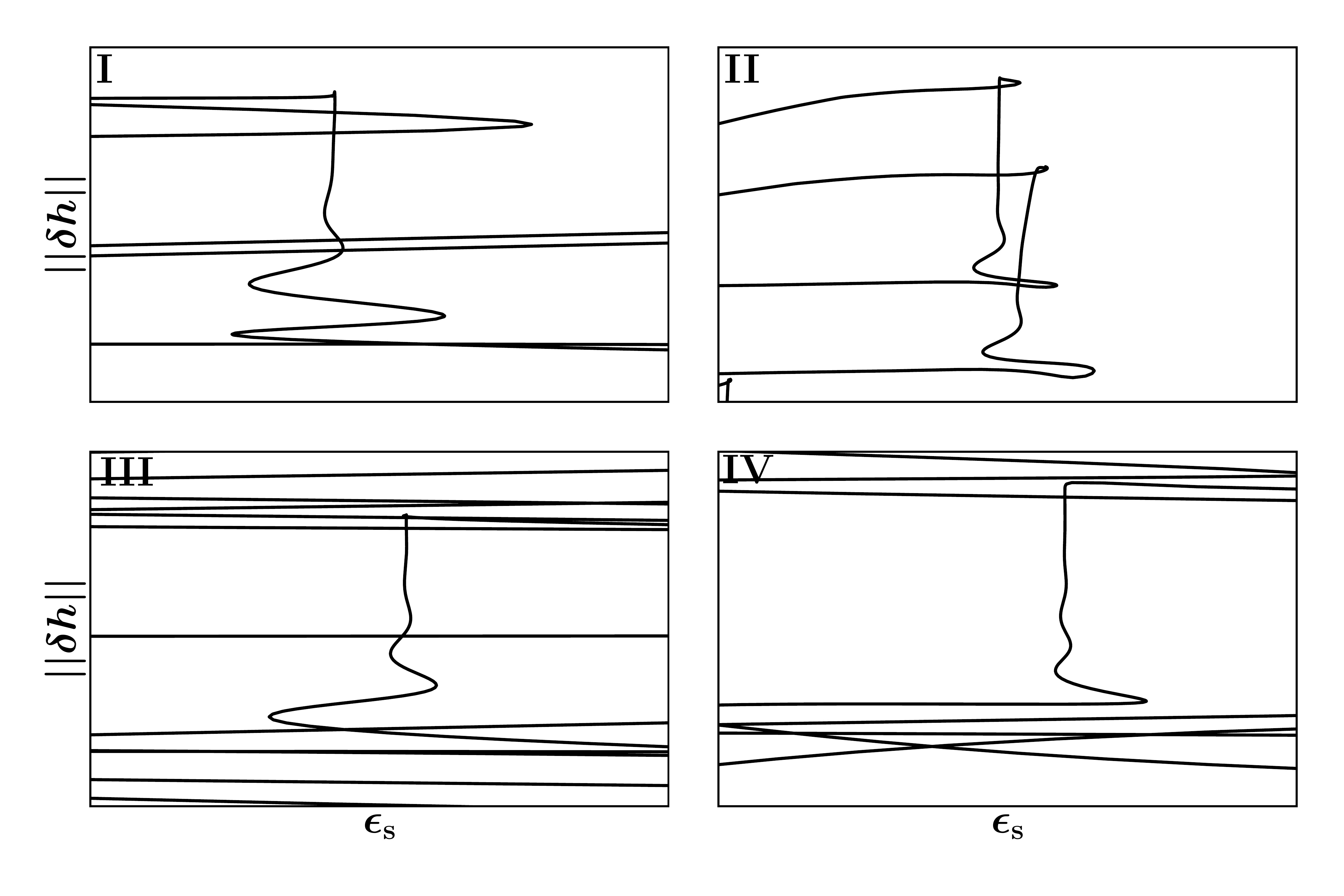}
\caption{Four zooms are given for the regions highlighted with red frames in Fig.~\ref{sec:mod:saw:fig:two_dimensional}(right) where exponential snaking structures occur. The data set can be found at \cite{LMTG2020zen}. 
}
\label{sec:mod:saw:fig:two_dimensional_zoom_further}
\end{figure*}

Next, we turn to the fully 2d system by transversally extending the domain to $L_x\times L_y = 40 \times 15$. The resulting bifurcation structures are given in the left panel of Fig.~\ref{sec:mod:saw:fig:two_dimensional} together with the 1d bifurcation curve already presented in Fig.~\ref{sec:mod:saw:fig:one_dimensional}. Overall the behaviour is similar as in the LB system:  At small $\epsilon_\mathrm{s}$ the 2d result coincides with the 1d curve, i.e., the solutions are $y-$translation invariant. Then, very closely after the saddle-node bifurcation of the 1d curve, a pitchfork bifurcation occurs where states with broken $y-$translation symmetry emerge. The emerging branch first extends towards smaller $\epsilon_\mathrm{s}$, undergoes a small number of saddle-node bifurcations (becoming linearly stable for a small range). Then after a large excursion it closely approaches the curve of 1d states where it wildly criss-crosses back and forth producing an extremely dense cluster of sub-branches. A zoom into the region indicated by the blue rectangular frame is given in Fig.~\ref{sec:mod:saw:fig:two_dimensional}(right). Solutions of the 2d branches in this region are all unstable. The branch did not reconnect to the 1d branch in our calculation. 
Carefully inspecting the seemingly erratic wiggling curve in Fig.~\ref{sec:mod:saw:fig:two_dimensional}(right) one discerns certain regular repeating structures that consist of a few saddle-node bifurcations ending in a short piece of vertical line. Several of these structures are highlighted by red frames and corresponding zooms are given in Fig.~\ref{sec:mod:saw:fig:two_dimensional_zoom_further}. As parts of the 1d curve and larger structures of the 2d curve they all closely resemble exponential snaking \cite{MaBK2010pd} as discussed for such systems in \cite{GTLT2014prl,TsGT2014epje}.

\begin{figure*}[p]
  \centering
  \includegraphics[width=0.9\hsize]{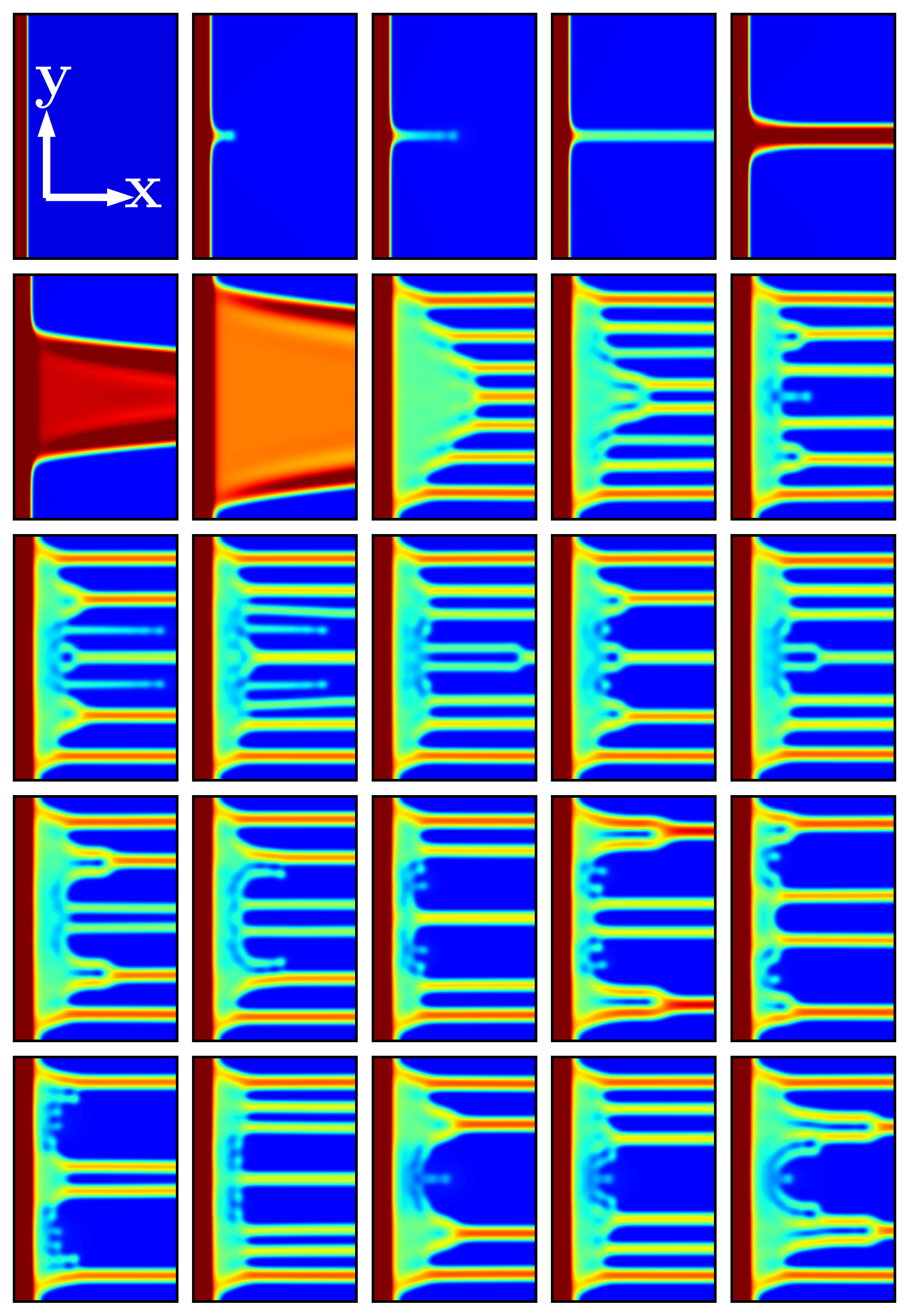}
\caption{Shown are selected film height profiles (for the corresponding colour bar see Fig.~\ref{sec:mod:saw:fig:two_dimensional_arclength_epsilon_y} of 2d steady states on the branch of fully 2d states in Fig.~\ref{sec:mod:saw:fig:two_dimensional}. From top left to bottom right we start at small $\epsilon_\mathrm{s}$ first follow the branch of transversally invariant states, switch to the branch of 2d states and follow it through to the end of our computation, i.e.,  we follow the branch with increasing arclength $s$. All states except panels 1 and 4 on the first row are unstable. Note that patterns are shown in the $y-$periodic $L_x\times 2L_y=40\times30$ domain. For better visibility the $y-$axis scale is stretched by a factor 2 as compared to one of the $x-$axis. The remaining parameters are as in Fig.~\ref{sec:mod:saw:fig:two_dimensional}.
}
\label{sec:mod:saw:fig:two_dimensional_steady_states}
\end{figure*}

\begin{figure*}[p]
\includegraphics[width=\hsize]{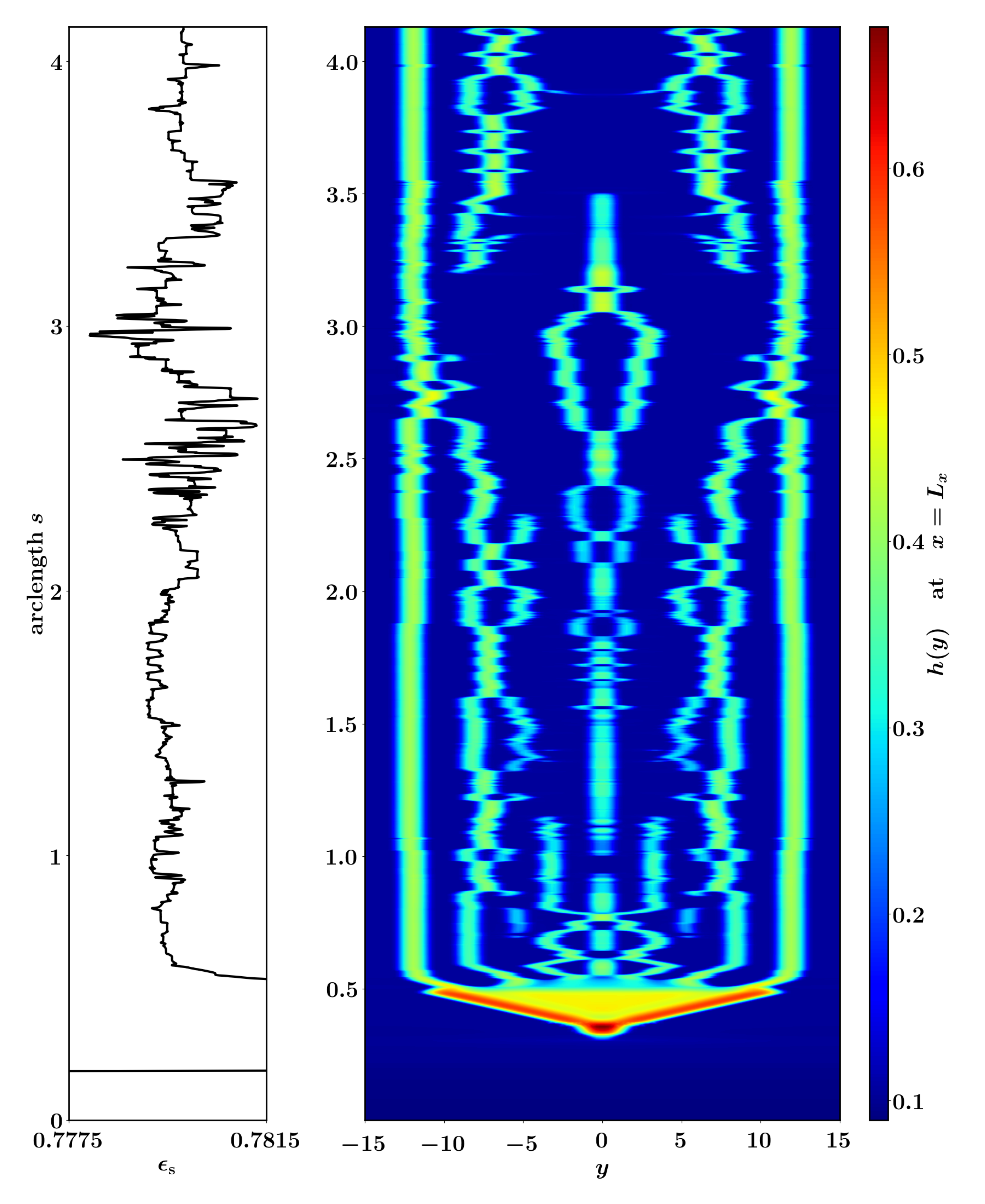}
\caption{(right) Shown are space-arclength plots showing the film height profile $h(x=L_x, y)$, i.e., on the downstream boundary, for a range of arclengths $s$ following the bifurcation curve of 2d states shown in Fig.~\ref{sec:mod:saw:fig:two_dimensional} through to the end of our computation. (left) The accompanying panel shows the corresponding values of the control parameter $\epsilon_\mathrm{s}$. The remaining parameters are as in Fig.~\ref{sec:mod:saw:fig:two_dimensional}.
}
\label{sec:mod:saw:fig:two_dimensional_arclength_epsilon_y}
\end{figure*}

Fig.~\ref{sec:mod:saw:fig:two_dimensional_steady_states} presents a small selection of 2d film height profiles along the solution branch shown in Fig.~\ref{sec:mod:saw:fig:two_dimensional}. Here, blue and red represent small and large film height, respectively. Note that calculations are performed using a $L_x\times L_y$ computational domain, while the images show a complete transversal period using the reflection symmetry at $y = 0$, i.e., they show a  $L_x\times 2L_y$ domain. In the sequence of the images we first follow the branch of transversally invariant states (panel~1 in first row), switch to the emerging branch of 2d states and follow it towards smaller $\epsilon_\mathrm{s}$ to its first fold. We see a central finger prolong to the right till it meets the right boundary (panels~2-4 in first row). A part of the branch with a sequence of saddle-node bifurcations in form similar to exponential snaking occurs centred about $\epsilon_\mathrm{s} \approx 0.845$ (this is a very small feature not well visible in the figure). Note that because of the finite system size in $y-$direction one is able to move between different branches that for a semi-infinite system would all remain disconnected.

Coming back to the bifurcation curve at hand we note that the single-finger state becomes stable (panel~4 on the first row of Fig.~\ref{sec:mod:saw:fig:two_dimensional_steady_states}) at another saddle-node bifurcation. The state stays stable until $||\delta h \approx 1.52||$, where it destabilises. 
The branch then continues in a rather long nearly vertical part. In parallel the finger structure widens till nearly filling the entire $y-$span, it flattens and finally consists of a broad plateau with higher rims at its lateral edges. After the vertical part that rises up to $||\delta h|| \approx 1.7$, the curve bends over and continues to smaller norm and smaller $\epsilon_\mathrm{s}$. 

When the norm strongly decreases, the very wide central finger decays into seven smaller fingers with a spacing that slightly increases from the center to the lateral boundaries (panel~3 in second row of Fig.~\ref{sec:mod:saw:fig:two_dimensional_steady_states}. Then one enters the wild and seemingly erratic region focused on in Fig.~\ref{sec:mod:saw:fig:two_dimensional}(right). However, what takes place there is actually a ``dance of fingers''  (panel~4 in second row till final panel): fingers break up into two smaller fingers, fingers fuse, fingers get shorter receding from the downstream boundary or get longer approaching and reaching the boundary. Thereby, points where fingers connect may move left or right resulting in rootlike structures visible in the images (e.g., in the final images of rows 2 to 4). It resembles a closing or opening zipper. The stripe number moves between one at the beginning to seven and then can move down to three and up to nine. This is very well visible in the space-arclength plot given in Fig.~\ref{sec:mod:saw:fig:two_dimensional_arclength_epsilon_y}.

In the seemingly erratic part, whose calculation we broke off at some point, the curve shows certain repeating elements. We highlight four of them by the red frames in Fig.~\ref{sec:mod:saw:fig:two_dimensional}(right) and provide the corresponding zooms in Fig.~\ref{sec:mod:saw:fig:two_dimensional_zoom_further}. There is always a short exponential snaking part followed by an approach to a vertical asymptote. For each of these pieces a steady state structure moves towards the right boundary or away from it, e.g., a finger increases or decreases in length or a point where a finger splits into two or two fingers join into one moves left or right. The top end of the vertical line, i.e., where it bends downwards again (seemingly horizontal parts in the extreme zooms) corresponds to the point when the structure reaches or leaves the right boundary. If a domain with larger [smaller] $x-$extension $L_x$ is studied the vertical pieces become longer [shorter]. In a system infinite in $x-$direction, the individual curves would disconnect. 

The visualisation as space-arclength plot given in Fig.~\ref{sec:mod:saw:fig:two_dimensional_arclength_epsilon_y} helps to understand what changes occur in the ``complex'' part of the bifurcation diagram. The dance of the fingers becomes more transparent when comparing the information in the adjusted left and right panel, i.e., the height profile on the downstream boundary and the value $\epsilon_\mathrm{s}$ both as a function of the arclength $s$. Each turn in the left panel corresponds to a saddle-node bifurcation in Fig.~\ref{sec:mod:saw:fig:two_dimensional}(right). Although, we do not analyse this plot in detail we highlight a number of interesting features:
Once created, the two outermost fingers remain in existence nearly unchanged for the full continuation, only the lateral position oscillates a bit and at about $s\approx2.8$ another respective finger fuses into them followed by to split-up and subsequent fuse events. The central finger has a more interesting life: it frequently disappears when splitting into two fingers and reappears when two fingers fuse, in two instances it disappears 'into the blue' (at $s\approx0.92$ and $s\approx3.5$) while in two other instances it equally suddenly appears (at $s\approx0.32$ and $s\approx1.02$). In the region $1.7 \le s \le 1.9$ a region of repeating oscillations between five and six fingers can be seen.

%%%%%%%%%%%%%%%%%%%%%%%%%%%%%%%%%%%%%%%%%%%%%%%%%%%%%%%%%%%%%%%%%%%%%%%%%%%%%%%%%
\section{Conclusion} \label{sec:conclusion}
%%%%%%%%%%%%%%%%%%%%%%%%%%%%%%%%%%%%%%%%%%%%%%%%%%%%%%%%%%%%%%%%%%%%%%%%%%%%%%%%%
 %
We have presented a brief investigation of the bifurcation structure related to the formation of two-dimensional deposition patterns as described by continuum models of Cahn-Hilliard type. In particular, we have considered a driven Cahn-Hilliard model for Langmuir-Blodgett transfer on the one hand and a driven thin-film description of a SAW-driven meniscus on the other hand. In both cases, we have presented and discussed bifurcation diagrams for one-dimensional and two-dimensional structures and selected corresponding steady states profiles. In the two-dimensional case we have given the profiles together with representations as space-arclength plots that allow one to appreciate morphology changes occurring along complicated 'criss-crossing' bifurcation curves.

Comparing both driven systems - Langmuir-Blodgett transfer and SAW-driven meniscus - we find that the overall scheme is qualitatively very similar when considering the dependency on the respective driving strength, namely, substrate velocity $v$ in Eq.~\eqref{eq:lb} and SAW strength $\epsilon_s$ in Eq.~\eqref{sec:mod:saw:eq:saw}.
In the one-dimensional case, starting from small driving strength, both systems exhibit stable homogeneous 'deposition', corresponding to the meniscus coexisting with a ultrathin adsorption layer (SAW driving) or homogeneous depositions of the LC phase (LB transfer). In both cases, this linearly stable one-dimensional state looses its stability in a saddle-node bifurcation, that is followed by a snaking-like part of the bifurcation curve related to the adjoinment of undulations to the front between bath concentration and deposited LC phase (LB transfer) or to the prolongation of an undulated foot structure (SAW driving). At large values of the control parameter, stable homogeneous deposition emerges that is stabilized by a Hopf bifurcation. A further increase in driving strength monotonically changes the value of the deposited field. For SAW driving this corresponds to Landau-Levich film transfer.

When considering two-dimensional domains, we have found an amended bifurcation scenario: In both cases, starting from small driving strength, a transversally invariant state is linearly stable that corresponds up to numerical deviations to the respective one-dimensional result. Then, close to the original first saddle-node bifurcation a pitchfork bifurcation occurs where a branch emerges that consists of states with broken transversal translation symmetry, i.e., transversal front modulations and ultimately finger-like structures develop.

Details differ between the two systems as in the LB system a second linearly stable one-dimensional state is approached before a further (this time imperfect) pitchfork bifurcation occurs that leads to the development of finger structures spanning the entire streamwise domain.
In contrast, in the SAW-driven system, the central finger directly grows from an initial bump without the intermediate stage of a laterally growing bump. When the finger reaches the downstream end of the domain the finger state stabilises and remains stable in a small range of driving strength. In contrast, in the LB system no stable finger states are found at the present parameters.

Beyond the discussed stage, in both cases the single bifurcation curve rather wildly criss-crosses back and forth thereby producing a dense cluster of sub-branches. The structure seems less erratic in the LB case than in the SAW case most likely because the typical lateral structure length compared to the lateral domain size is smaller in the SAW case than in the LB case, i.e., the number of 'combinatoric possibilities' is much larger in the former case. Another difference is the occurrence of circular spots in the  LB case that do nearly not occur in the SAW case. Vice versa the moving zipper structures are prevalent in the SAW case but do not occur in the LB case. A systematic investigation of these finer details and their dependence on parameters has to be left open for future research. This is also the case for time-periodic states that we have left out entirely. For both systems a large number of Hopf bifurcations has been detected on the branch of fully two-dimensional states. It remains a formidable numerical challenge for the future to follow and analyse the emerging two-dimensional time-periodic states. We believe that up to now for systems of the considered type this has only been done for one-dimensional states \cite{KoTh2014n,LRTT2016pf,TWGT2019prf}.

In general, our results on the bifurcation structure of fully two-dimensional states in driven Cahn-Hilliard-type systems should be relevant for a large number of systems involving coating processes, advancing liquid fronts or receding liquid fronts. For instance, transversal instabilities are often described for fronts of liquid that move down an inclined plate \cite{VeIC1998jfm}. There, the formation of fingers is frequent, is normally studied by linear stability analysis and/or direct time simulations \cite{Hupp1982n,Schw1989pfa,SpHo1996pf,Kall2000jfm,ErSR2000pf,ThKn2003pf,KoDi2004pf}, and could strongly benefit from the application of continuation techniques. 

Regrettably, nearly all of the studied transversally structured states have turned out to be unstable. Therefore a major future aim should be to identify conditions that render some of these or related structures stable and to better understand the time scales and detailed dynamics of the time evolutions when starting at slightly perturbed unstable states. Further, we point out that unstable steady states are sometimes more easily stabilised by external forcing than other (arbitrary) states. This can, e.g., be done by prestructured substrates \cite{GHLL1999s,KaSh2001prl,KoDi2002pre,KoDi2004pf,WiGu2014pre,HLHT2015l,ZWHH2016ami,WZCT2017jpcm} or spatio-temporally structured forcing of the inflow at the bath or meniscus \cite{LTCG2019pre}. A detailed understanding of the aimed at unstable states could then inform how to construct the external forcing.

%%%%%%%%%%%%%%%%%%%%%%%%%%%%%%%%%%%%%%%%%%%%%%%%%%%%%%%%%%%%%%%%%%%%%%%%%%%%%%%%%

\section*{Acknowledgements}
We acknowledge support by
the Deutsche Forschungsgemeinschaft (DFG; Grant No.~TH781/8-1 and PAK 943 with Grant No.~332704749), % Coating, LB
and the German-Israeli Foundation for Scientific Research and Development (GIF, Grant No. I-1361-401.10/2016).
We acknowledge discussions with Roman Kleine-Arndt regarding the numerical path continuation for the LB system in the two-dimensional case and Sebastian Engelnkemper for corresponding preliminary work. 
We thank Arik Yochelis and the other organisers for arranging the workshop \textit{Advances in pattern formation: New questions motivated by applications} (Sede Boqer, Feb.\ 2019) in honour of Ehud Meron’s 65th birthday and all participants for the lively and inspiring discussions.

\section*{Bibliography}
\bibliography{dipCoatingReview}

\end{document}